\documentclass[aps,twocolumn,prd,longbibliography]{revtex4-1}    
\pdfoutput=1
\usepackage{bm}
\usepackage{graphicx}
\usepackage{booktabs}
\usepackage{slashed}
\usepackage{hyperref}
\usepackage{amssymb,amsmath,amsfonts}
\usepackage{color}
\usepackage[utf8]{inputenc}
\usepackage[caption=false]{subfig}
\usepackage[utf8]{inputenc}
\usepackage[caption=false]{subfig}

\begin{document}

\title{Roman Jackiw and Chern-Simons theories}

\author{Robert D. Pisarski}
\affiliation{Department of Physics, Brookhaven National Laboratory, Upton, NY 11973, USA}
\begin{abstract}
  I recount my personal experience interacting with Roman Jackiw in the 1980's, when we both worked on Chern-Simons theories  in three dimensions.
\end{abstract}

\maketitle

Recently, the Center of Mathematical Sciences and Applications at Harvard initiated an excellent series of talks on mathematics.
In the inaugural talk on March 13, 2020,
S.-T. Yau spoke about S. S. Chern as a great geometer of the 20th century  \cite{yau}.  Particularly in the discussion section after the talk,
Prof. Yau emphasized the essential role
which Roman Jackiw played in bringing Chern-Simons theories into physics.  In this note I wish to share some recollections from
those times, and especially my interactions with Roman.   I do this because I most strongly agree with Prof. Yau's assessment of Roman's contribution.
In this as other areas, Roman's work exhibits both the sheer joy of computation, while
continually pushing to understand the greater significance of what is truly new and exciting.

As a graduate student of David Gross, with Larry Yaffe we computed the fluctuations to one loop order
about a single instanton at nonzero temperature \cite{Gross:1980br}.
I then went off to Yale University as a postdoc \cite{yale}, where I worked with Tom Appelquist on gauge theories in three dimensions.
Our motivation was to understand the behavior of gauge theories at high temperature, which for
the static mode reduces to a gauge theory in three dimensions.  We computed the gluon self energy to one and (in part) to two loop order.
The gluon self energy isn't gauge invariant, so the greater significance of our analysis wasn't clear.  But we also computed in an Abelian
theory for a large number of scalars, which is a nice soluble model, and from Tom I learned the most useful craft of power counting diagrams.  
We concentrated on the gluons, since quarks are fermions, and decouple in the static limit.

Sometime in the fall of 1980, I went to MIT to give a talk on this work.  It was a joint seminar with Harvard, and I remember it well to this day.
Talks were then on transparencies, and the evening before I thought I would be clever, and added a comment that while
topological charge in four dimensions
is quantized classically, perhaps it isn't quantum mechanically.  Sidney Coleman was sitting near the front, in a purple crushed velvet suit
which to me looked very much like that of Superfly.  When he saw that slide, however, Coleman pounced, and did not let up until
I surrendered abjectly, admitting to the idiocy of my suggestion.  

During the talk and for the entire afternoon after, Roman grilled me about details of the calculation, how gauge theories in three dimensions work,
what about gauge invariance, everything.
It was very intense and quite exhilarating.  Roman and S. Templeton then wrote a paper on theories in three dimensions \cite{Jackiw:1980kv},
which because Roman is a superb calculator, appeared as a preprint a few weeks before ours.

When their paper came out, I remember looking at it, and thinking, ah, very good, they were spending their time on something
irrelevant, two component fermions in three dimensions.  Now if you forget about nonzero temperature and just do dimensional reduction,
the natural thing is to go from four component fermions in four dimensions to four component fermions in three dimensions.

And thus I missed the really new physics, which Roman grasped.  The most interesting part of the paper by Roman and Templeton
appears a bit pedestrian, at the beginning of Sec. III, Eqs. (3.3) and (3.4).  
Under the Lorentz (or Euclidean) group in three dimensions, all one needs are two component fermions,
since in three dimensions one can just take the Dirac matrices to be the Pauli matrices.
What I did not work out is that under the discrete transformations of parity and charge conjugation,
that a mass term for a {\it single} two component fermion is parity {\it odd}!  Roman and Templeton
discuss in their Ref. (11) \cite{Jackiw:1980kv}: the dimensional reduction
of a four component fermion gives {\it two} two-component fermions.  The masses for these are of equal in magnitude,
but {\it opposite} in sign.  This is how the mass for four dimensional fermions, which is certainly parity even, remains so
after dimensional reduction to three dimensions.

Thus while if wasn't present in the original paper by Roman and Templeton, if one computes the gauge self energy for {\it two}-component
fermions in three dimensions, then a parity mass term for the gauge field will appear {\it immediately}.  Yes, it has nothing
to do with nonzero temperature, but so what?  It is an absolutely beautiful, novel, and gauge invariant mass term, special to three dimensions.

This was first proposed in two papers by S. Deser, Roman, and Templeton, in a Physical Review Letter  \cite{Deser:1982vy},
and a long paper in Annals of Physics \cite{Deser:1981wh}.  The year before, Jonathan Schonfeld has proposed the same theory
\cite{Schonfeld:1980kb}.
What Deser, Roman, and Templeton realized, however, and which Schonfeld did not, is that there is something extraordinary about a non-Abelian
Chern-Simons term \cite{cs}:  the ratio of the Chern-Simons mass term to the gauge coupling is an integer,
\begin{equation}
  q = \frac{4 \pi m}{g^2} = {\rm integer} \; ,
\end{equation}
where $m$ is the Chern-Simons mass, and $g^2$ is the coupling constant for the non-Abelian gauge theory.
This quantization of the topological ratio $q$ follows from invariance under large and topologically nontrivial gauge transformations, as the Chern-Simons
term is related to the topological charge in four dimensions.

For years I didn't believe this result; it just seemed so {\it simple}, how could it possibly be right?
For a while I went off in other directions.  When I was at Fermilab a few years later, though, with Sumathi Rao we thought that if
one computes perturbatively, as an expansion in $g^2/m$, then to one loop order the result will be a pure number,
independent of $g^2$ or $m$.  And {\it surely} perturbation theory can't know about large gauge transformations, right?

Sumathi and I did the computation to one loop order by brute force, which was not trivial.  We had to sort out
differences with the previous results of Deser, Roman, and Templeton \cite{Deser:1981wh}.

In the end, by computing a ratio of (finite) renormalization constants, we discovered that in a SU(N) gauge theory the topological ratio shifts as
\begin{equation}
  q \rightarrow q + N \; ,
\end{equation}
That is, even in perturbation theory $q$ shifts precisely by the number of colors, and thus
knows about large gauge transformations!  This is an example of how in physics, one can do good things for all the wrong
reasons.  I set out to prove that Roman was wrong, and instead, completely confirmed his results in a beautiful and unexpected way.

I then continued working on Chern-Simons theories, considering the effects at nonzero temperature \cite{Pisarski:1986gq} and the effects of
magnetic monopoles \cite{Pisarski:1986gr}.
I showed that the topological ratio remains quantized at nonzero temperature \cite{uhlenbeck}, and that 
even in the Abelian theory, magnetic monopoles can produce the quantization of the topological ratio.

However, this was the time of the Second String Revolution, and my work had little impact upon the field.  Except for Roman, who would include me
in correspondence to others working in the field, such as Andrei Linde and Oleg Kalashnikov.  I include copies of these letters in the figures below,
from the time of our original work.  The most
important quote is from Roman, which unfortunately is partially obscured by a marker I used on the original.  He wrote:

{\it The gauge-invariant mass term that we have discovered may in fact provide the correct infra-red regulation at high temperatures.  At present
  we do not see how it can be derived from the four-dimensional high-temperature theory: perhaps it has its origin in the $\theta F \widetilde{F}$
  term.  I would be very interested in any comments you may have about this.}

Even now I do not understand how a Chern-Simons mass term in three dimensions can arise from a $\theta$ term in four dimensions, but that is really 
besides the point.  Roman's work illustrates the virtue of following things where they lead.
Don't solve the problem you want to solve, solve the problem waiting to be solved.

In 1988, Edward Witten did his magesterial work on the relationship between Chern-Simons theories and the Jones polynomial \cite{Witten:1988hf}.
Along the way, almost as an aside, he derived the shift in the topological ratio $q$.
The work by Rao and I had so little impact upon the field that Alvarez-Baume, Labastida, and Ramallo
\cite{AlvarezGaume:1989wk}, and also Chen, Semenoff, and Wu \cite{Chen:1989tk}, rederived this result.
In writing this article, I was rather amused to see that the shift in the topological ratio $q$ was later derived by Axelrod and Singer
using perturbative means \cite{Axelrod:1991vq}.

I close with a personal comment.  After Fermilab I went to the High Energy Theory Group at Brookhaven.  Years after I was hired, Bill Marciano mentioned
to me once that Roman had most strongly supported my hire.  Of course this wasn't the only support I had: surely kind words from my advisor David Gross,
from Tom Appelquist, and from Larry McLerran \cite{mcl}, amongst others, also helped.
Nevertheless, I was most struck by Bill's comment.  In a very real sense, a good part of
why I have been blessed to pursue theoretical physics,
this bizarre and ahuman activity, is due to Roman.   This note is my way of thanking him.

\acknowledgments
This research was supported by the U.S. Department of Energy 
under contract DE-SC0012704.

\begin{figure}[t]
  \centering
 \includegraphics[scale=0.9]{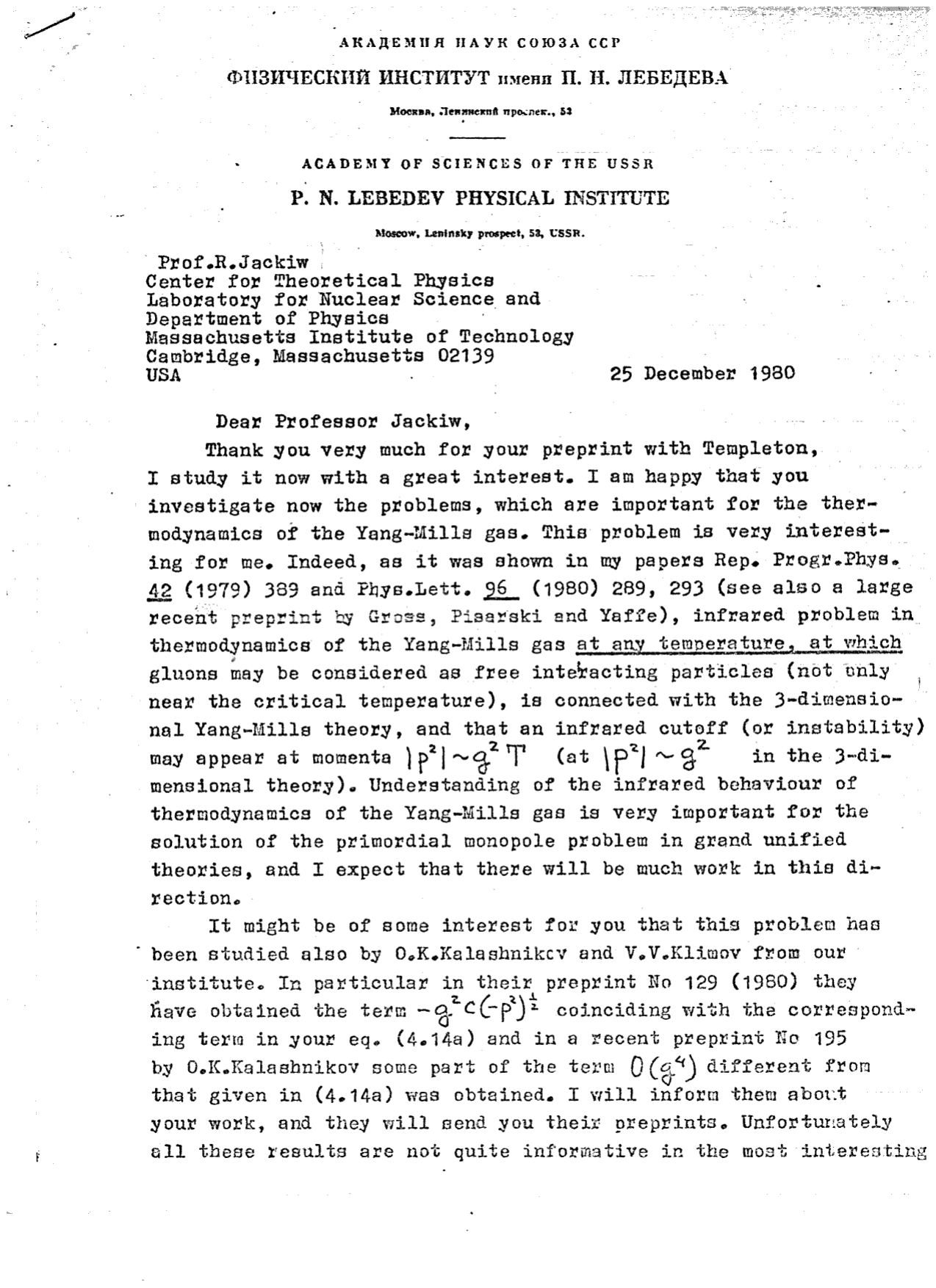}
  \caption{Letter from A. Linde to Roman, page 1.}
\end{figure}

\begin{figure}[t]
  \centering
  \includegraphics[scale=0.9]{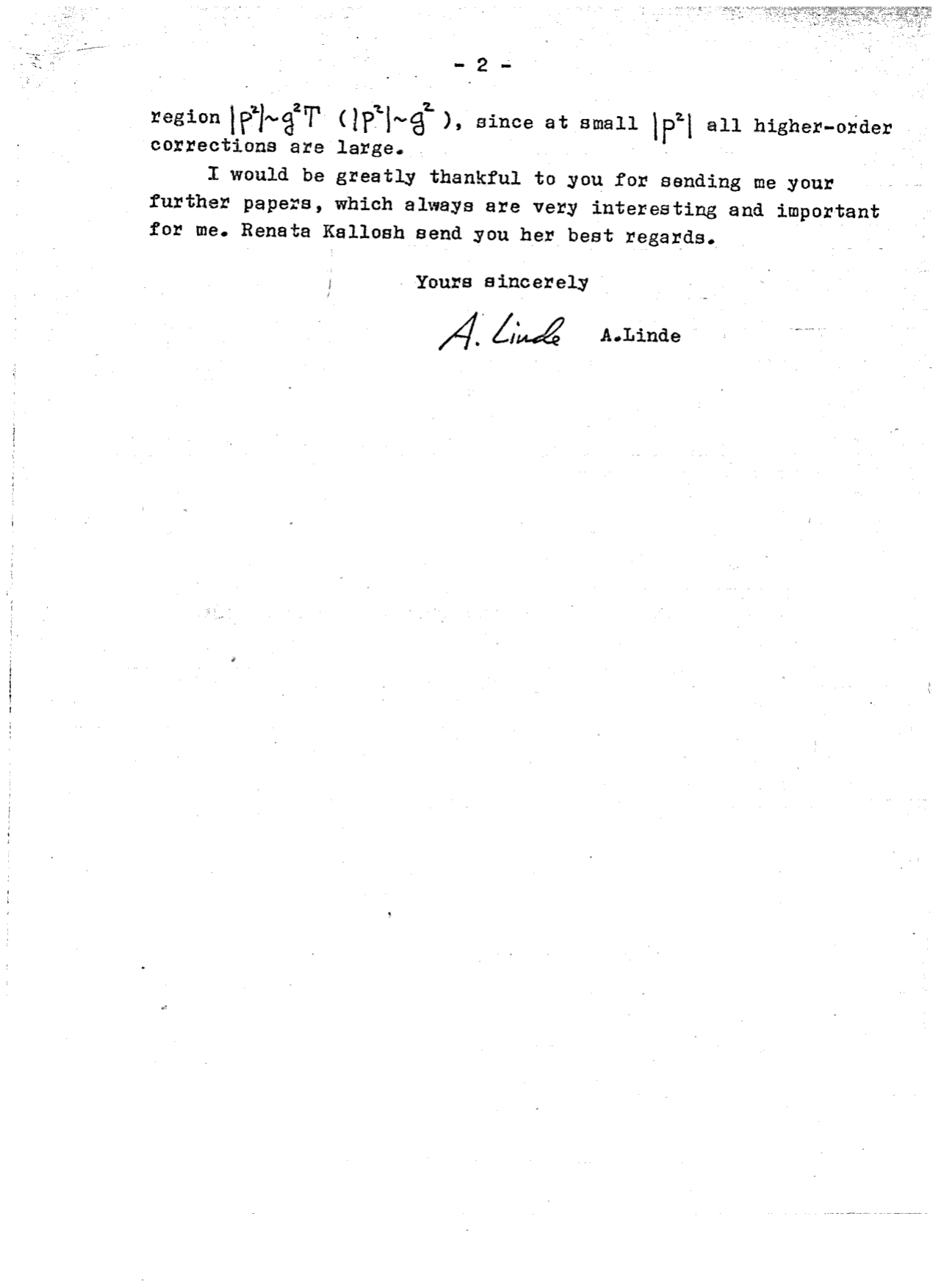}
  \caption{Letter from A. Linde to Roman, page 2.}
\end{figure}

\begin{figure}[t]
  \centering
\includegraphics[scale=0.9]{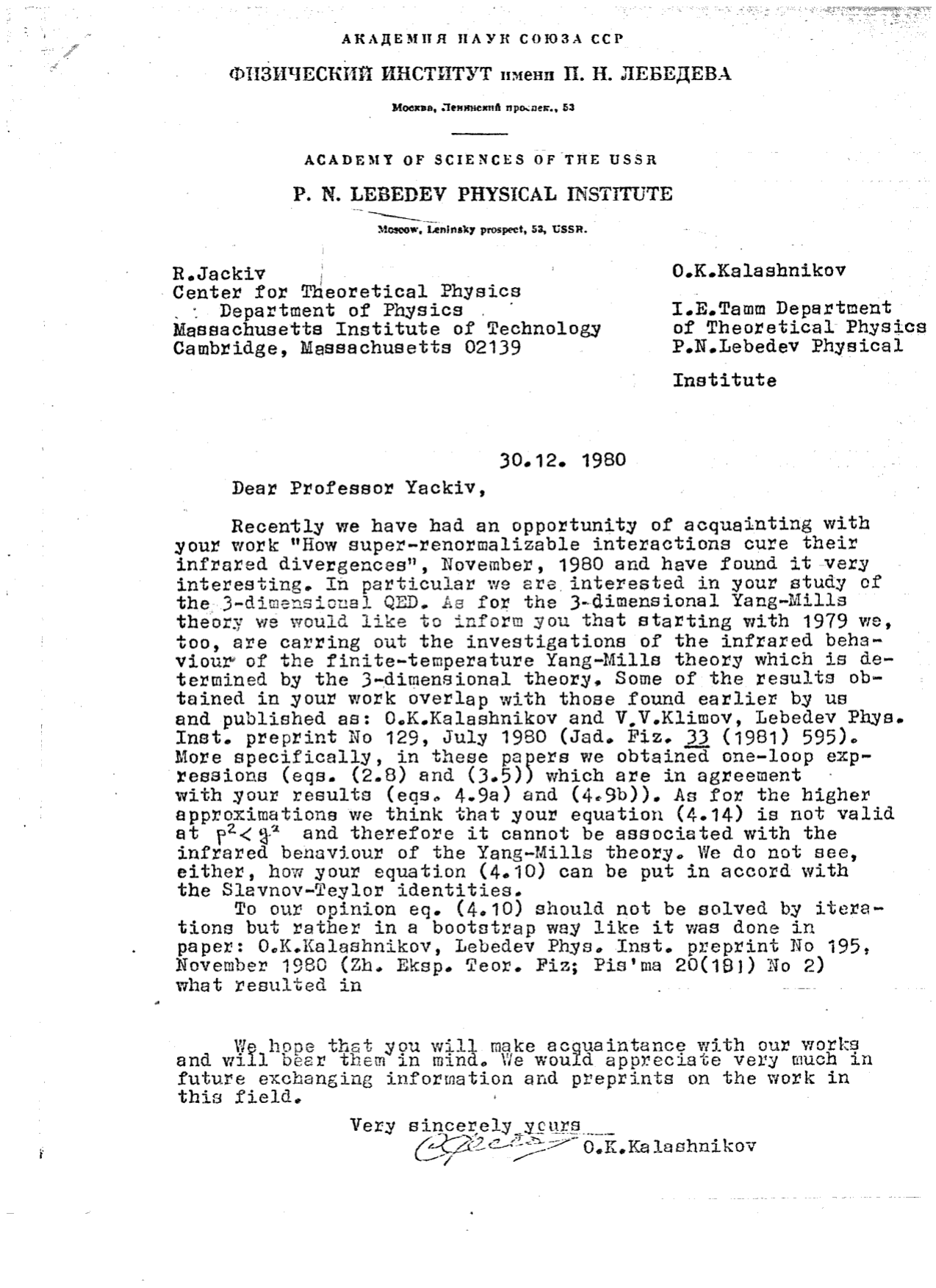}
  \caption{Letter from O. Kalashnikov to Roman.}
\end{figure}

\begin{figure}[t]
  \centering
\includegraphics[scale=0.9]{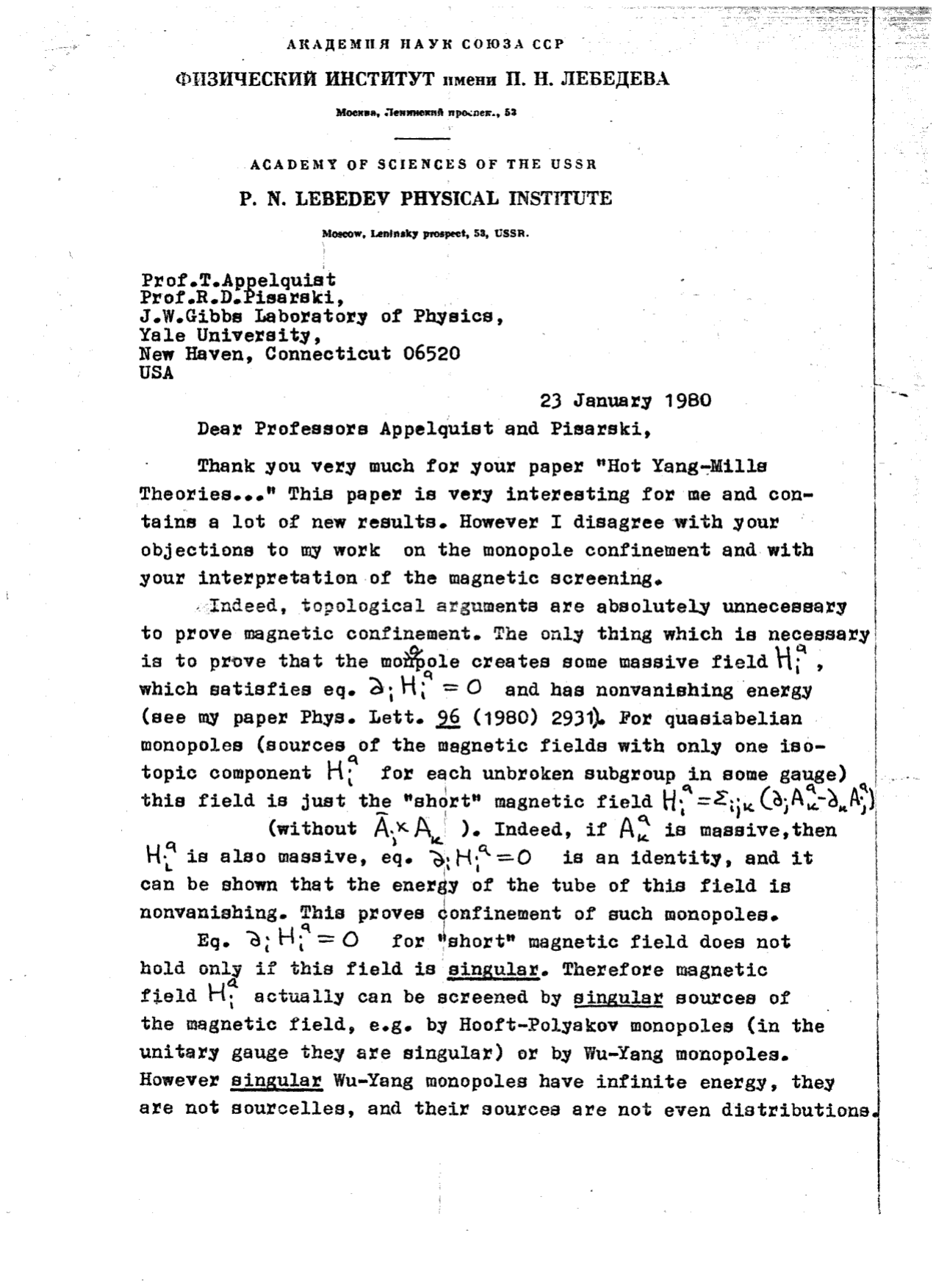}
  \caption{Letter from A. Linde to T. Appelquist and myself, page 1.}
\end{figure}

\begin{figure}[t]
  \centering
  \includegraphics[scale=0.9]{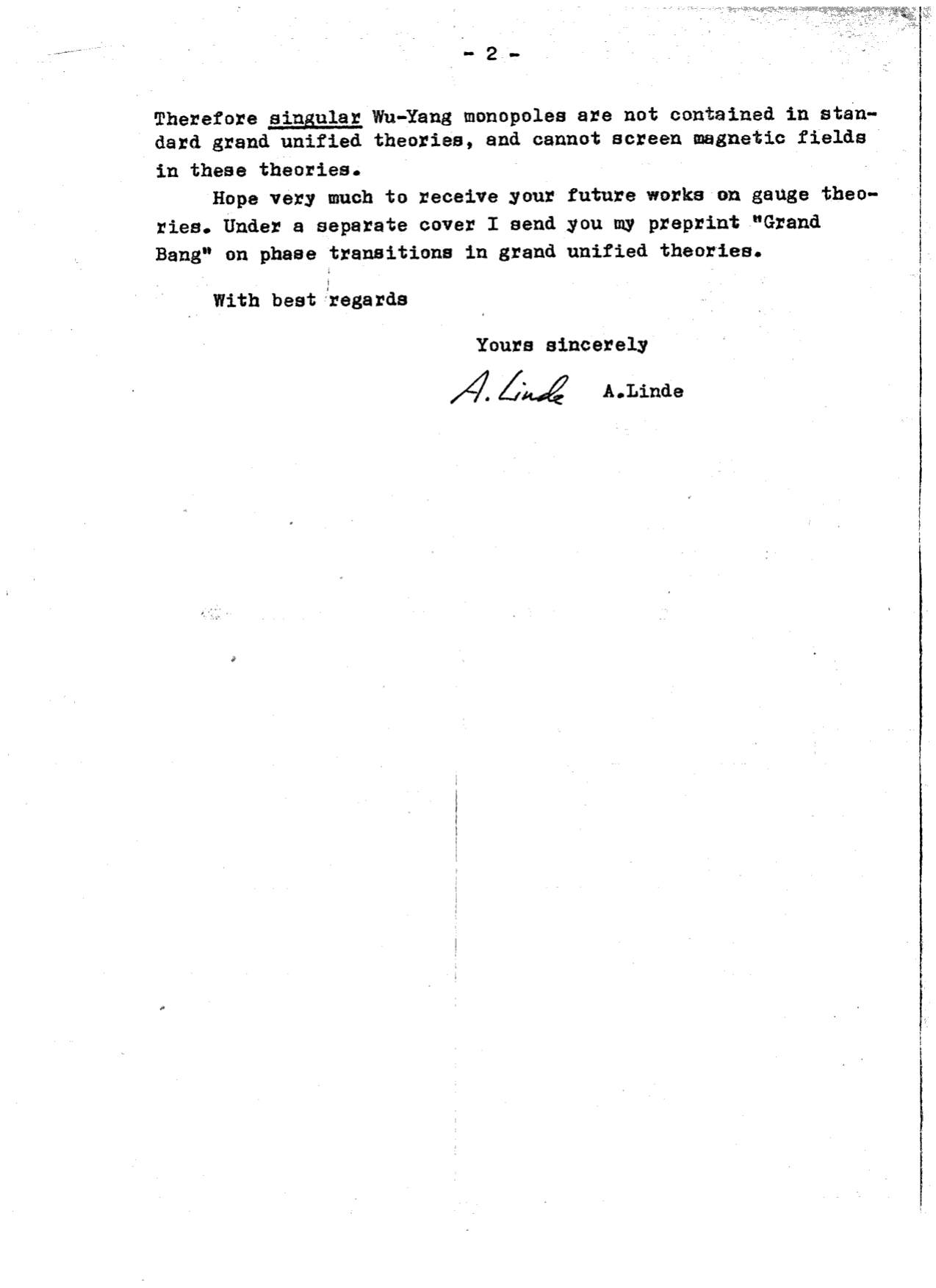}
  \caption{Letter from A. Linde to T. Appelquist and myself, page 2.}
\end{figure}

\begin{figure}[t]
  \centering
\includegraphics[scale=0.9]{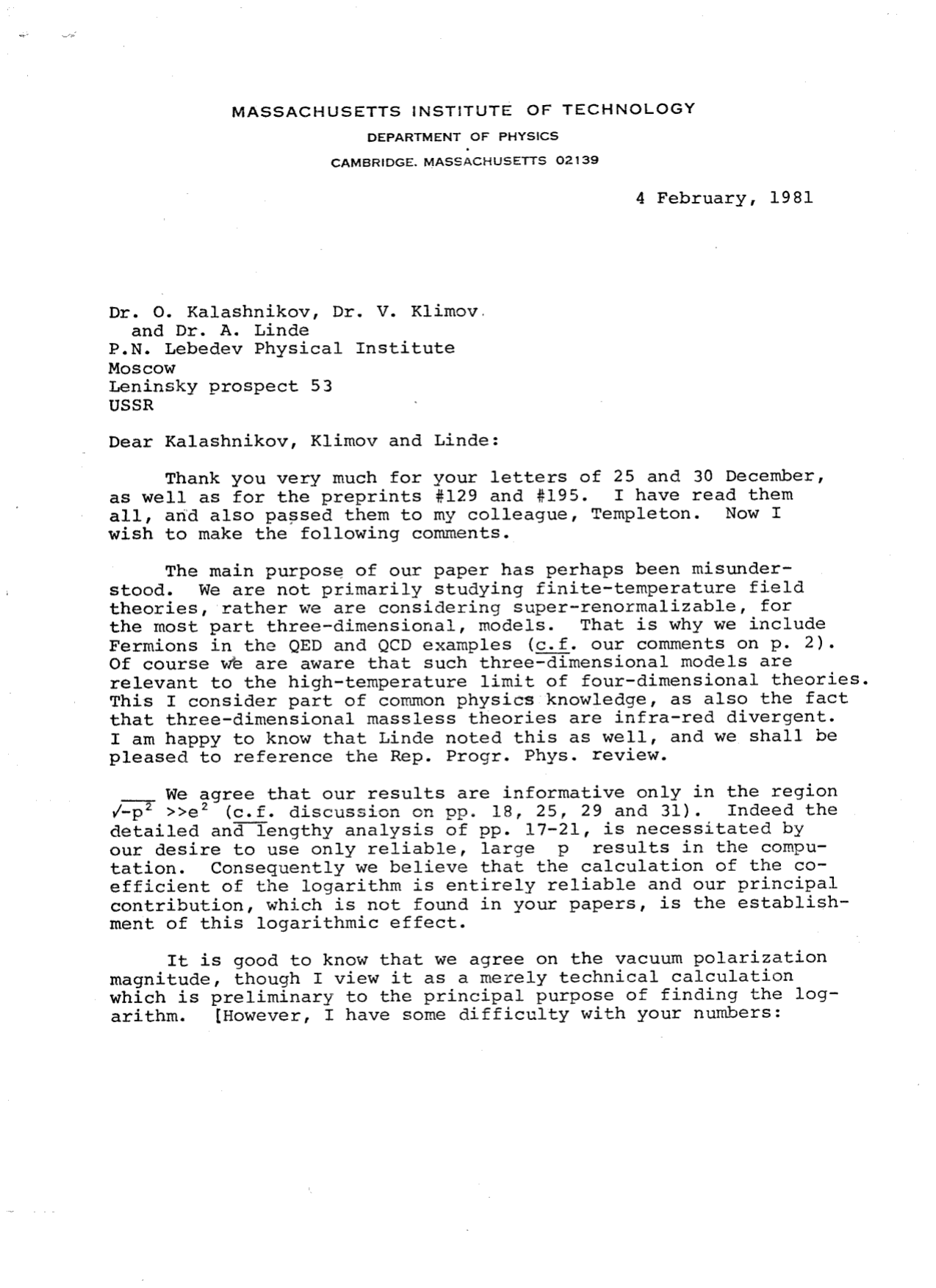}
  \caption{Letter from Roman to A. Linde and O. Kalashnikov, page 1.}
\end{figure}

\begin{figure}[t]
  \centering
\includegraphics[scale=0.9]{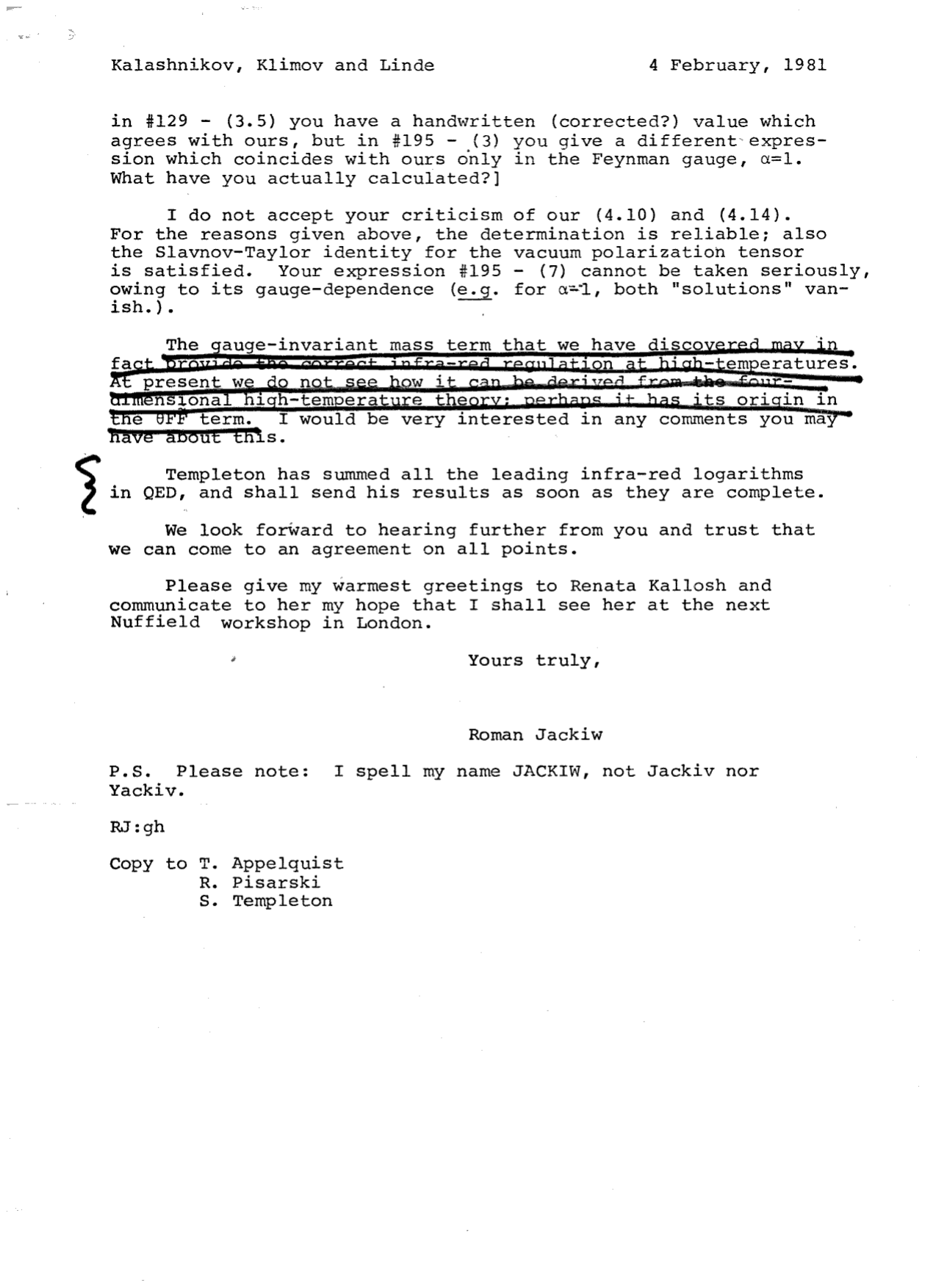}
  \caption{Letter from Roman to A. Linde and O. Kalashnikov, page 2.}
\end{figure}


\begin{thebibliography}{9}
\bibitem{yale}I was a J. W. Gibbs Fellow, in fact the last.  Since much of my career after working on Chern-Simons theories was devoted to the statistical mechanics
of non-Abelian gauge theories, I am especially proud to have been associated, albeit tangentially, with the first great American theoretical physicist.
\bibitem{cs}Back then we used the term topologically massive, but here I refer to the mass term as Chern-Simons, since they did invent it first.
\bibitem{mcl}While at Fermilab, Larry and Peter Arnold developed their analysis of sphaleron transitions, and I spent many hours arguing with them, convinced that
they were wrong.    As with the quantization of the topological charge, again the error was mine.
\end{thebibliography}
\end{document}